\documentclass[aps,twocolumn,showpacs]{revtex4}

\usepackage{graphicx}
\usepackage{dcolumn}
\usepackage{amsmath}
\usepackage[utf8]{inputenc}

\begin{document}
\title{Spin frustration and fermionic entanglement in an exactly solved hybrid diamond chain with the localized Ising spins and mobile electrons}
\author{J. Torrico$^1$, M. Rojas$^2$, M.~S.~S. Pereira$^1$, J. Stre$\check{c}$ka$^3$ and  M.~L.~Lyra$^1$}

\affiliation{$^1$Instituto de Física, Universidade Federal de Alagoas, 57072-970 Maceió, AL, Brazil}
\affiliation{$^2$Departamento de Física, Universidade Federal de Lavras, 37200-000, Lavras,MG, Brazil}
\affiliation{$^3$Department of Theoretical Physics and Astrophysics, Faculty of Science, P. J. $\check{S}$afárik University, Park Angelinum 9, 040 01 Ko$\check{s}$ice, Slovak Republic}

\begin{abstract}
The strongly correlated spin-electron system on a diamond chain containing localized Ising spins on its nodal lattice sites and mobile electrons on its interstitial sites is exactly solved in a magnetic field using the transfer-matrix method. We have investigated in detail all available ground states, the magnetization processes, the spin-spin correlation functions around an elementary plaquette, fermionic quantum concurrence and spin frustration. It is shown that the fermionic entanglement between mobile electrons hopping on interstitial sites and the kinetically-induced spin frustration are closely related yet independent phenomena. In the ground state, quantum entanglement only appears within a frustrated unsaturated paramagnetic phase, while thermal fluctuations can promote some degree of quantum entanglement above the non-frustrated ground states with saturated paramagnetic or classical ferrimagnetic spin arrangements.  
\end{abstract}
\maketitle

\section{Introduction}

The entanglement of physical states unveils the presence of non-local correlations in quantum spin systems that have no classical counterpart \cite{bell,amico,toth}.
Recently, the theoretical and experimental study of quantum entanglement has been put forward by a possibility of information processing at the quantum level \cite{nielsen}. The generation and manipulation of entangled states are fundamental in quantum information processes such as quantum computation \cite{bennett1}, teleportation \cite{bennett2,yeo}, cloning of quantum states \cite{chiara} and quantum cryptography \cite{ekert}. Along this direction, condensed matter systems play a central role in the study of quantum entanglement, because several quantum devices have been proposed on the basis of solid-state systems. It has been demonstrated that quantum entanglement can affect the low-temperature behavior of macroscopic properties, such as the magnetic susceptibility and specific heat \cite{ghosh,vedral,brukner,vertesi}. This fact has raised the interest in exploring the relation between quantum entanglement and thermodynamic macroscopic observables \cite{souza}. 

Within the above scenario, quantum spin chains serve as ideal model systems in studies of entanglement signatures \cite{connor,imamoglu,rappoport}. In particular, it has been shown that quantum entanglement exhibits a characteristic scaling behavior in a vicinity of the quantum critical point present in a class of one-dimensional quantum spin systems \cite{osterloh}. This feature has opened a new perspective for the description of quantum phase transitions \cite{verstraete,verstraete2,popp} and resulted in the introduction of new quantities to measure the degree of quantum entanglement such as the quantum concurrence \cite{wootters,hillwootters}.

The quantum Heisenberg spin chain is the prototype model used to search for entanglement signatures in the thermodynamic properties of magnetic systems \cite{arnesen,wang1,wang2,kamt,connor,wang3}. In an isotropic Heisenberg system composed of two qubits, the quantum concurrence decreases as a function of temperature until it completely vanishes at a characteristic threshold temperature, which depends on the applied external magnetic field. Further, the degree of quantum entanglement decreases monotonically with increasing the magnetic field for all temperatures \cite{arnesen}. On the other hand, Starace et. al. \cite{kamt} showed that the anisotropy and external field can be used to produce and control in a two-qubit XY Heisenberg chain the degree of quantum entanglement at any finite temperature. 

In the last two decades, quantum diamond spin chains have been also subject to extensive investigations. The spin-1/2 Heisenberg diamond chain \cite{takano,okamoto,wang4,honecker,gu,mikeska,rule,kang,ananikian} as well as its simplified Ising-Heisenberg version \cite{canova,lisni,lisny,valverde} have been largely explored in connection with a possible interplay between geometric spin frustration and quantum fluctuations. These systems display a rich variety of unusual physical properties such as magnetization plateaus \cite{rule,canova,lisni,lisny}, doubled peaks in the specific heat and susceptibility \cite{gu,kang} and quantum entanglement \cite{ananikian}. In addition, magnetic properties of A$_{3}$Cu$_{3}$(PO$_{4}$)$_{4}$ with A=Ca,Sr \cite{drillon1,drillon2}, Cu$_{3}$(TeO$_{3}$)$_{2}$Br$_{2}$ \cite{uematsu}, K$_3$Cu$_3$AlO$_2$(SO$_4$)$_4$ \cite{fujihala}, Cu$_{3}$(CO$_{3}$)$_{2}$(OH)$_{2}$ (azurite) \cite{kikuchi} and Cu$_3$(OH)$_5$(NO$_3$) (likasite) \cite{kikuchin} are satisfactorily captured by 
various versions of the spin-1/2 Heisenberg diamond chain. High-field magnetization measurements on the natural minerals azurite and likasite have for instance confirmed an existence of one-third \cite{kikuchi} and two-thirds \cite{kikuchin} magnetization plateau besides the double peaks in the relevant thermodynamic response functions. 

A presence of the intermediate magnetization plateau has also been reported for the hybrid diamond-chain model, in which the spins of mobile electrons delocalized over decorating quantum dimers (interstitial sites) interact with each other as well as with the localized Ising spins situated at nodal lattice sites \cite{pereira1,pereira2,lisni1,lisni2,nalbandyan}. The hybrid spin-electron diamond chain with localized Ising spins and delocalized electrons has been exactly solved using an exact diagonalization procedure in combination with the decoration-iteration mapping technique. In this class of models, the competition emerges from the quantum-mechanical hopping of the interstitial electrons and is termed as the kinetically-driven spin frustration \cite{pereira1,pereira2}. 

While the quantum entanglement has been rather extensively studied in the Ising-Heisenberg diamond chains \cite{ananikian1,rojas1,torrico}, it has been almost untouched in the analogous spin-electron diamond chains. It has been evidenced for the spin-1/2 Ising-Heisenberg diamond chain that the ground state is disentangled (classical) when the Ising coupling between the nodal and interstitial spins is the predominant one, but it becomes entangled for a strong enough Heisenberg coupling between the interstitial spins. The quantum concurrence generally decreases due to thermal fluctuations until it completely vanishes above a certain threshold temperature. The dependence of the quantum concurrence on the temperature and external field was explored for example in the Ising-XXZ \cite{rojas1} as well as Ising-XYZ \cite{torrico} diamond chain. In the latter model, the XY anisotropy may result in a re-entrant behavior of the quantum entanglement. 

Although thermodynamic properties of the hybrid spin-electron diamond chain have been already explored in some detail \cite{pereira1,pereira2,lisni1,lisni2,nalbandyan}, the inter-relation between the kinetically-driven spin frustration and the fermionic quantum entanglement is still missing\cite{kozlowski}. In this work, we will address this question by considering an exactly solvable diamond chain with localized Ising spins and mobile electrons as a prototype model. The kinetic term associated with the electron mobility generates competition and spin frustration. We will report the ground state phase diagram including ferrimagnetic (FRI), saturated paramagnetic (SPA) and unsaturated paramagnetic (UPA) phases. We will examine in detail the magnetization process, which exhibits an intermediate plateau, as well as the spin correlations around the plaquette. A frustrated regime will be characterized by the absence of a local order that simultaneously satisfy all first-neighbors correlations. We will also compute the 
fermionic quantum concurrence between a pair of interstitial electrons in order to evaluate the influence of thermal fluctuations in the degree of quantum entanglement and its relation with spin frustration. We will unveil distinct regimes of quantum entanglement as a function of the hopping amplitude and the external field, including the emergence of a re-entrant fermionic concurrence in the vicinity of the zero-temperature FRI-UPA and SPA-UPA ground-state phase transitions. 

This work is organized as follows: In section II, we will describe the  model Hamiltonian and present the methodology used to obtain the exact solution for the magnetization, correlation functions and quantum concurrence. In section II, we present the ground state phase diagram, and a detailed study of the magnetization processes, correlation functions, spin frustration and quantum entanglement.  In section IV, we summarize and draw our main conclusions.  Some details of the analytical derivation are given in the appendices.

\section{Model and its Hamiltonian}

\begin{figure}[htbp]
\centering
\includegraphics [scale=0.4,clip]{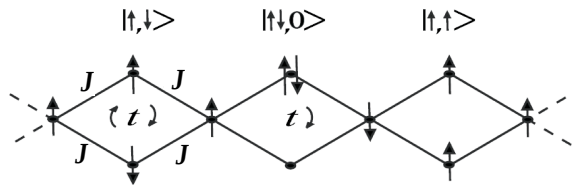}
\caption{A diagrammatic representation of a correlated spin-electron model on a diamond chain. The nodal sites are occupied by the localized Ising spins $\sigma_i = \pm 1$, which are coupled by the Ising interaction with the spin of mobile electrons hopping between the nearest-neighbor interstitial sites.}
\label{H1}
\end{figure}

Let us consider a coupled spin-electron model on a diamond chain, which contains localized Ising spins $\sigma_i=\pm 1$ on its nodal sites and two mobile electrons
on each couple of interstitial sites (see Fig. \ref{H1}). For further convenience, the total Hamiltonian can be defined as a sum over cell Hamiltonians, 
i.e. ${\cal H} = \sum_{i} {\cal H}_i$, whereas each cell Hamiltonian ${\cal H}_i$ involves all the interaction terms belonging to $i$-th diamond unit:
\begin{eqnarray}
{\cal H}_i \!\!\!&=&\!\!\! - t \sum_{\gamma=\uparrow,\downarrow} \left(c_{i1,\gamma}^{\dagger} c_{i2,\gamma} + {\rm h.c.}\right)
               - h \sum_{j=1}^2 \left( n_{ij,\uparrow} - n_{ij,\downarrow} \right) \nonumber \\
        \!\!\!&+&\!\!\! J \left(\sigma_i + \sigma_{i+1}\right)\sum_{j=1}^2 \left( n_{ij,\uparrow} - n_{ij,\downarrow} \right) \nonumber \\
               \!\!\!&-&\!\!\! \frac{h}{2} \left(\sigma_i + \sigma_{i+1}\right)\!.
\label{ham}
\end{eqnarray}
Here, $c_{ij,\gamma}^{\dagger}$ and $c_{ij,\gamma}$ $(j = 1,2)$ are the usual fermionic creation and annihilation operators for the mobile electrons with the spin $\gamma= \uparrow$ or $\downarrow$ and $n_{ij} =  c_{ij,\gamma}^{\dagger} c_{ij,\gamma}$ is the respective number operator. The hopping term $t$ takes into account the  kinetic energy related to a quantum-mechanical hopping of the mobile electrons on the interstitial sites, the coupling constant $J$ accounts for the nearest-neighbor Ising interaction between the localized Ising spins and the mobile electrons,  and $h$ is Zeeman's energy of the localized Ising spins and mobile electrons in a presence of the external magnetic field.

The matrix form of the Hamiltonian (\ref{ham}) in the local basis of two mobile electrons for the $i$-th diamond cell $\left|\uparrow,\uparrow\right>_i$, $\left|\downarrow,\downarrow\right>_i$, $\left|\uparrow\downarrow,0\right>_i$, $\left|\uparrow,\downarrow\right>_i$, $\left|0,\uparrow\downarrow\right>_i$ and $\left|\downarrow,\uparrow\right>_i$ can be represented as:
\begin{eqnarray}
{\cal H}_i = \left( \begin{array}{cccccc}
 2J\mu_i-2h & 0 \rule{3mm}{0mm}& 0 \rule{3mm}{0mm}& 0 \rule{3mm}{0mm}& 0 \rule{3mm}{0mm}& 0 \\ \\
 0 & -2J\mu_i+2h \rule{3mm}{0mm}& 0 \rule{3mm}{0mm}& 0 \rule{3mm}{0mm}& 0 \rule{3mm}{0mm}& 0 \\ \\
 0 & 0 \rule{3mm}{0mm}& 0 \rule{3mm}{0mm}& t \rule{3mm}{0mm}& 0 \rule{3mm}{0mm}& t\\ \\
 0 & 0 \rule{3mm}{0mm}& t \rule{3mm}{0mm}& 0 \rule{3mm}{0mm}& t \rule{3mm}{0mm}& 0\\ \\
 0 & 0 \rule{3mm}{0mm}& 0 \rule{3mm}{0mm}& t \rule{3mm}{0mm}& 0 \rule{3mm}{0mm}& t\\ \\
 0 & 0 \rule{3mm}{0mm}& t \rule{3mm}{0mm}& 0 \rule{3mm}{0mm}& t \rule{3mm}{0mm}& 0 \\
\end{array}\right)  \label{6}
\end{eqnarray}
where we have introduced the notation for the total spin of two Ising spins from $i$-th diamond unit $\mu_i=\sigma_i+\sigma_{i+1}$. For simplicity, we have left out the constant term $-h \mu_i/2$ as it can be later simply added to the corresponding eigenvalues of the Hamiltonian matrix (\ref{6}). A straightforward diagonalization of the Hamiltonian matrix (\ref{6}) leads (after taking into account also the constant term $-h \mu_i/2$) to the following spectrum of eigenvalues:
\begin{eqnarray}
\varepsilon_{i1}&=&2J\mu_i-2h-\frac{h}{2}\mu_i\label{7},\\
\varepsilon_{i2}&=&-2J\mu_i+2h-\frac{h}{2}\mu_i\label{8},\\
\varepsilon_{i3}&=&-\frac{h}{2}\mu_i+2t\label{9},\\
\varepsilon_{i4}&=&-\frac{h}{2}\mu_i-2t\label{10},\\
\varepsilon_{i5}&=&-\frac{h}{2}\mu_i \label{11},\\
\varepsilon_{i6}&=& -\frac{h}{2}\mu_i \label{12},
\end{eqnarray}
which correspond to the eigenvectors
\begin{eqnarray}
\left|\varphi_{i1}\right>&=&\left|\uparrow,\uparrow\right>_i\label{13},\\
\left|\varphi_{i2}\right>&=&\left|\downarrow,\downarrow\right>_i\label{14},\\
\left|\varphi_{i3}\right>&=&\frac{1}{2}[\left|\downarrow,\uparrow\right>_i+\left|\uparrow,\downarrow\right>_i+\left|0,\uparrow\downarrow\right>_i+\left|\uparrow\downarrow,0\right>_i] \label{15},\\
\left|\varphi_{i4}\right>&=&\frac{1}{2}[\left|\downarrow,\uparrow\right>_i+\left|\uparrow,\downarrow\right>_i-\left|0,\uparrow\downarrow\right>_i-\left|\uparrow\downarrow,0\right>_i] \label{16},\\
\left|\varphi_{i5}\right>&=& \frac{1}{\sqrt{2}}[\left|0,\uparrow\downarrow\right>_i-\left|\uparrow\downarrow,0\right>_i]\label{17},\\
\left|\varphi_{i6}\right>&=&  \frac{1}{\sqrt{2}}[\left|\downarrow,\uparrow\right>_i-\left|\uparrow,\downarrow\right>_i]\label{18}.
\end{eqnarray}
All physical quantities of interest can be subsequently calculated from these exact analytical results by adapting the standard transfer-matrix method.

\subsection{Magnetization and correlation functions}

Here, we will derive exact results for the magnetization and correlation functions, as provided by the transfer-matrix technique. The partition function of the correlated spin-electron model on a diamond chain can be calculated following the procedure:
\begin{eqnarray}
\mathcal{Z}_N&=&\sum_{\{\sigma \}} \prod_{i=1}^{N} \mbox{Tr}_i e^{-\beta {\cal H}_i}
            = \sum_{\{\sigma \}} \prod_{i=1}^{N} \sum_{j=1}^6 e^{-\beta \varepsilon_{ij}} \nonumber \\
           &=&\sum_{\{\sigma \}} \prod_{i=1}^{N} \omega(\sigma_i, \sigma_{i+1}) = \mbox{Tr} \, W^N = \lambda_{+}^N + \lambda_{-}^N,
\label{19}
\end{eqnarray}
where $\beta=1/(k_B T)$, $k_B$ is Boltzmann's constant, $T$ is the absolute temperature, the summation $\sum_{\{\sigma \}}$ is carried out over all possible spin configurations of the localized Ising spins, the symbol $\mbox{Tr}_i$ refers to the trace over the degrees of freedom of two mobile electrons from the $i$-th diamond unit and $W = \omega(\sigma_i, \sigma_{i+1})= \sum_{j=1}^6 e^{-\beta \varepsilon_{ij}}$. The expression $W$ can be regarded as the transfer matrix:
\begin{eqnarray}
W&=&\left( \begin{array}{clrr}
\omega(1,1) & \omega(1,-1) \\
\omega(-1,1) & \omega(-1,-1)
\end{array}\right) \nonumber \\
&=&\left( \begin{array}{clrr}
\omega(2) & \omega(0) \\
\omega(0) & \omega(-2)
\end{array}\right),
\label{20}
\end{eqnarray}
whereas individual matrix elements depend just on the total spin of two localized Ising spins from $i$-th diamond unit according to:
\begin{eqnarray}
\omega(\mu_i) &=& \omega(\sigma_i,\sigma_{i+1}) =\sum_{j=1}^6 \exp(-\beta \varepsilon_{ij}) = 2 \exp\left( \frac{\beta h}{2} \mu_i \right) \nonumber \\
             &\times&  \left[1 + \cosh (2 \beta t) + \cosh (2 \beta J \mu_i - 2 \beta h) \right].
\label{bw}
\end{eqnarray}
The exact result for the partition function (\ref{19}) then readily follows from the two eigenvalues of the transfer matrix (\ref{20}):
\begin{eqnarray}
\lambda_\pm= \frac{1}{2} \left[\omega(2)+\omega(-2) \pm Q\right]
\label{21},
\end{eqnarray}
where $Q=\left([\omega(2)-\omega(-2)]^2+4[\omega(0)]^2\right)^\frac{1}{2}$. In the thermodynamic limit of $N\rightarrow\infty$, only the largest eigenvalue effectively contributes to the partition function.

The unitary transformation that diagonalizes the transfer matrix (\ref{20}), is determined by the matrices $U$ and $U^{-1}$:
\begin{eqnarray}
\textbf{U}=\left( \begin{array}{clrr}
\frac{\lambda_+-\omega(-2)}{A_+} & \frac{\lambda_--\omega(-2)}{A_-}\\
\frac{\omega(0)}{A_+} & \frac{\omega(0)}{A_-}\\
\end{array}\right) \label{22}
\end{eqnarray}
where $A_{\pm}= \sqrt{[\lambda_{\pm}-\omega(-2)]^2 + [\omega(0)]^2}$ and $U^{-1}=U^{\dagger}$.

Now, one may take advantage of the unitary transformation in order to calculate the magnetization and correlation functions. For instance, the magnetization of localized Ising spins at the nodal lattice sites follows from the relation:
\begin{eqnarray}
M_\sigma=\left<\sigma_i^z\right>=\frac{1}{\mathcal{Z}_N} \mbox{Tr} \left[\tilde{\sigma}^z\tilde{W}^{N}\right],
\label{27}
\end{eqnarray}
where $\tilde{\sigma}^z=U^{-1}\sigma^zU$ and $\tilde{W}=U^{-1}W U$. $\sigma^z$ is the standard $2\times 2$ Pauli matrix associated with the z-component of a nodal Ising spin. A similar formula can also be derived for the
magnetization at the interstitial site $j=1,2$ of the $i$-th cell:
\begin{eqnarray}
M_S=\left<S_{ij}^z\right>=\frac{1}{\mathcal{Z}_N} \mbox{Tr} \left[\tilde{\Sigma}_{ij}^z\tilde{W}^{N-1}\right],
\label{26}
\end{eqnarray}
where $\tilde{\Sigma}_{ij}^z = U^{-1} \Sigma_{ij}^z U$, with the elements $\Sigma_{ij}^z(\sigma_i,\sigma_{i+1})=\mbox{Tr}~S_{ij}^ze^{-\beta H_i}$ and the trace performed over the states of the interstitial sites belonging to the $i$-th cell.  The matrix $S_{ij}^{z}$ corresponds to the $z$-component of the spin operator at the interstitial site $j=1,2$ of the $i$-th cell, written in the corresponding $6\times 6$ sub-space.  The total magnetization can be then obtained either by adding both individual contributions of the localized nodal Ising spins (\ref{27}) and the mobile electrons (\ref{26}), or respectively, from the usual formula:
\begin{eqnarray}
M=-\frac{1}{\beta}\frac{\partial}{\partial h}\ln{{\cal Z}_N}.
\label{28}
\end{eqnarray}
The same procedure can be used to calculate the correlation functions. The correlation functions between various spatial components of spin operators $S_{ij}^{\alpha}$ at the interstitial sites of an elementary plaquette can be calculated as \cite{bellucci}:
\begin{eqnarray}
\left<S_{i1}^\alpha S_{i2}^\alpha\right>=\frac{1}{\mathcal{Z}} \mbox{Tr} \left(\tilde{\Sigma}_{12}^{\alpha}\tilde{W}^{N-1}\right)
\label{24},
\end{eqnarray}
where $\tilde{\Sigma}_{12}^{\alpha}=U^{-1} \Sigma_{12}^{\alpha}U$, with the elements  $\Sigma_{12}^{\alpha}(\sigma_i,\sigma_{i+1})= \mbox{Tr}~  S_{i1}^\alpha S_{i2}^\alpha e^{-\beta H_i}$. The correlation function between the localized Ising spins and the interstitial spins is given by:
\begin{eqnarray}
\left<\sigma_i S_{ij}^z\right>=\frac{1}{\mathcal{Z}_N}\mbox{Tr}\left[\tilde{\sigma}^z\tilde{\Sigma}_{ij}^{z}\tilde{W}^{N-1}\right].
\label{25}
\end{eqnarray}

\subsection{Fermionic entanglement}

Quantum entanglement is closely related to non-local correlations present in a quantum system. For two qubits, it can be quantified by the  entanglement of formation \cite{wootters}:
\begin{eqnarray}
E_F=H\left(\frac{(1+\sqrt{1-\mathcal{C}^2})}{2}\right),
\label{29}
\end{eqnarray}
where
\begin{eqnarray}
H(x)=-x\log_2(x)-(1-x)\log_2(1-x),
\label{30}
\end{eqnarray}
and $C$ is called the quantum concurrence given by:
\begin{eqnarray}
\mathcal{C}=\rm{max}\left\{0, \sqrt{\Lambda_1}-\sqrt{\Lambda_2}-\sqrt{\Lambda_3}-\sqrt{\Lambda_4}\right\} \label{31}.
\end{eqnarray}
Here, $\Lambda_i$ are the eigenvalues of the matrix $R = \rho\left(\sigma^y\otimes\sigma^y\right)\rho^\ast\left(\sigma^y\otimes\sigma^y\right)$ sorted in descending order, $\sigma^y$ is the usual Pauli matrix and $\rho$ represents the reduced density matrix for a pair of qubits. 
Because the entanglement of formation (\ref{29}) is a monotonous function of the concurrence $C$, one may directly use the concurrence
as a measure of quantum entanglement ranging from 0 (no entanglement) up to 1 (maximum entanglement) \cite{wootters,hillwootters}. 

In the present model, we will quantify the quantum entanglement in the sub-space of up spins in a given plaquette \cite{shu}. Therefore, in the local basis $\left|0,0\right>$,  $\left|0,\uparrow\right>$,  $\left|\uparrow,0\right>$ and  $\left|\uparrow,\uparrow\right>$, the reduced density matrix can be written in the general form
\begin{eqnarray}
\rho_{i,j}=\left( \begin{array}{clrr}
a & 0 & 0 & 0 \\
0 & x & z & 0 \\
0 & z^* & y & 0 \\
0 & 0 & 0 & b
\end{array}\right) \label{41},
\end{eqnarray}
whose elements can be directly associated with the following  thermal averages
\begin{eqnarray}
b&=&\left<n_{i1,\uparrow},n_{i2,\uparrow}\right> \\ \nonumber
x&=&\left<n_{i1,\uparrow}\right>-\left<n_{i1,\uparrow},n_{i2,\uparrow}\right> \\ \nonumber
y&=&\left<n_{i1,\uparrow}\right>-\left<n_{i1,\uparrow},n_{i2,\uparrow}\right> \\ \nonumber
a&=&\left<(1-n_{i1,\uparrow})(1-n_{i2,\uparrow})\right> \\ \nonumber
z&=&z^*= \left<c_{i1,\uparrow}^\dag,c_{i2,\uparrow}\right>
\label{39}
\end{eqnarray}

Following the procedure elaborated previously in Ref. \cite{shu}, one arrives to the following explicit form of the eigenvalues:
\begin{eqnarray}
\Lambda_1 &=& \left(|\langle n_{i1,\uparrow} \rangle - \langle n_{i1,\uparrow} n_{i2,\uparrow} \rangle|
                + |\langle c_{i1,\uparrow}^{\dagger} c_{i2,\uparrow} \rangle| \right)^2, \nonumber \\
\Lambda_2 &=& \left(|\langle n_{i1,\uparrow} \rangle - \langle n_{i1,\uparrow} n_{i2,\uparrow} \rangle|
                - |\langle c_{i1,\uparrow}^{\dagger} c_{i2,\uparrow} \rangle| \right)^2, \nonumber \\
\Lambda_{3,4} &=& \langle n_{i1,\uparrow} n_{i2,\uparrow} \rangle \left(1 - 2 \langle n_{i1,\uparrow} \rangle
                + \langle n_{i1,\uparrow} n_{i2,\uparrow} \rangle \right).
\label{lamb}
\end{eqnarray}
According to Eqs. (\ref{lamb}), the fermionic concurrence can be calculated from the relation:
\begin{eqnarray}
\mathcal{C} = \!\!\!&2&\!\!\! {\rm max}\left\{ 0, |\langle c_{i1,\uparrow}^{\dagger} c_{i2,\uparrow} \rangle| \right. \nonumber \\ 
              \!\!\!&-&\!\!\! \left. \sqrt{\langle n_{i1,\uparrow} n_{i2,\uparrow} \rangle \left(1 - 2 \langle n_{i1,\uparrow} \rangle 
                + \langle n_{i1,\uparrow} n_{i2,\uparrow} \rangle \right)} \right\}, 
\label{Concur}
\end{eqnarray}
which depends on the statistical mean values $\langle c_{i1,\uparrow}^{\dagger} c_{i2,\uparrow} \rangle$, $\langle n_{i1,\uparrow} \rangle$ and $\langle n_{i1,\uparrow} n_{i2,\uparrow} \rangle$ explicitly given in Appendix B. It is noteworthy that the more general formula (\ref{Concur}) for the fermionic concurrence reduces to the simpler formula reported previously by Deng and Gu \cite{shu} when considering the particular case $\left< n_{i1,\uparrow}\right> =  \left< n_{i2,\uparrow}\right> = 1/2$.

\section{Results and discussions}

Now, let us proceed to a discussion of the most interesting results for the coupled spin-electron diamond chain. For simplicity, our further discussion will be restricted to a particular case with the antiferromagnetic Ising interaction $J>0$, which should exhibit the most diverse magnetic behavior on behalf of a mutual interplay between the kinetically-driven spin frustration, local quantum fluctuations and the Zeeman's splitting.

\subsection{Ground state}

\begin{figure}[htbp]
\centering
\includegraphics [scale=0.3,clip]{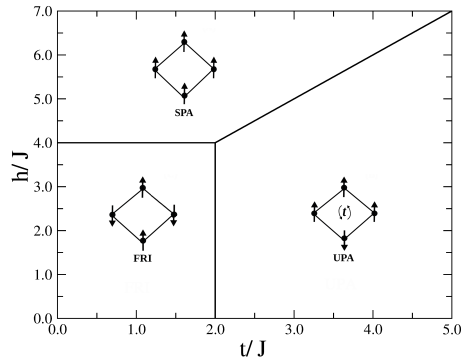}
\caption{The ground-state phase diagram in the $t/J$ vs. $h/J$ plane. In the UPA phase, the interstitial electrons are quantum entangled. It is highly degenerated at zero field because the nodal spins become effectively decoupled. A high degeneracy is also present at the triple point with the coordinates $t/J=2.0$ and $h/J=4.0$, where all three phases co-exist with a nodal antiferromagnetic phase (see Appendix A).}
\label{D1}
\end{figure}

First, we will turn our attention to the ground-state phase diagram shown in Fig. \ref{D1}, which involves the ferrimagnetic (FRI), the saturated paramagnetic (SPA) and the unsaturated paramagnetic (UPA) ground states unambiguously determined by the eigenvectors:
\begin{eqnarray}
\left|\rm{FRI}\right>&=&\prod_{i=1}^N\left|\varphi_{i1}\right>\otimes\left|\sigma_i=-1\right>,\nonumber \\
\left|\rm{SPA}\right>&=&\prod_{i=1}^N\left|\varphi_{i1}\right>\otimes\left|\sigma_i=1\right>, \nonumber \\
\left|\rm{UPA}\right>&=&\prod_{i=1}^N\left|\varphi_{i4}\right>\otimes\left|\sigma_i=1\right>\label{34},
\end{eqnarray}
with the corresponding eigenenergies per plaquette:
\begin{eqnarray}
E_{\rm{FRI}}&=&-4J-h,\nonumber \\
E_{\rm{SPA}}&=&4J-3h,\nonumber \\
E_{\rm{UPA}}&=&-h-2t\label{35}.
\end{eqnarray}
It can be understood from the eigenvectors (\ref{34}) that the interstitial spins are aligned into the magnetic field and the nodal spins in opposite to the magnetic field within the FRI ground state, while all nodal as well as interstitial spins are equally aligned into the magnetic field within the SPA ground state. However, the most peculiar spin arrangement can be found within the UPA ground state, where the nodal spins are aligned into the magnetic field but the interstitial spins are subject to a quantum entanglement of two antiferromagnetic and two ionic states as described by the eigenvector (\ref{16}). The phase UPA originates from a kinetically-driven spin frustration, which is closely connected to the quantum-mechanical hopping of two mobile electrons with opposite spins on the interstitial sites. In agreement with this statement, the nodal spins are completely free to flip in an absence of the magnetic field and the UPA ground state becomes highly degenerate. High macroscopic degeneracy can be 
also found at a triple point given by the coordinates $t/J=2.0$ and $h/J=4.0$, at which all available ground states FRI, SPA and UPA coexist together with another special nodal antiferromagnetic (NAF) ground state (see Appendix~A).

\subsection{Correlation functions and magnetization}

The local spin arrangements inherent to each individual ground state can be witnessed through the pair correlation functions, which can also give insights into how the spin-spin correlations are affected by thermal fluctuations.

\begin{figure}[htbp]
\centering
\includegraphics [scale=0.4,clip]{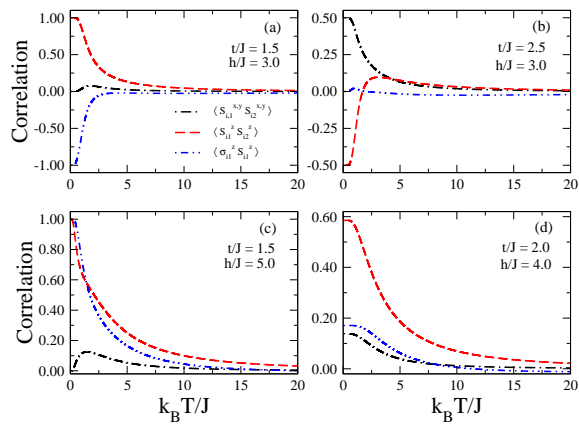}
\caption{Temperature variations of the transverse correlation between the interstitial spins $\left<S_{i1}^{x,y} S_{i2}^{x,y}\right>$, the longitudinal correlation between the interstitial spins $\left<S_{i1}^z S_{i2}^z\right>$, the longitudinal correlation between the interstitial and nodal spins $\left<\sigma_{i1}^z S_{i1}^z\right>$ for the set of parameters: (a) $t/J=1.5$, $h/J=3.0$ (FRI phase); (b) $t/J=2.5$, $h/J=3.0$ (UPA phase); (c) $t/J=1.5$, $h/J=5.0$ (SPA phase); (d) $t/J=2.0$, $h/J=4.0$ (triple point). Notice that spin frustration is present in the low temperature regime above the UPA ground state where only out of three longitudinal spin-spin correlations within the elementary plaquette is negative.}
\label{F1}
\end{figure}

In Fig.~\ref{F1} we have plotted, for all available ground states, typical thermal dependences of the transverse correlation function between the interstitial spins $\left<S_{i1}^x S_{i2}^x\right> = \left<S_{i1}^y S_{i2}^y\right>$ (dash-dot lines), the longitudinal correlation function between the interstitial spins $\left<S_{i1}^z S_{i2}^z\right>$ (dash lines) and the longitudinal correlation function between the nodal and interstitial spins $\left<\sigma_{i1}^z S_{i1}^z\right>$ (dash-dot-dot lines). The zero-temperature asymptotic limits of the correlation functions shown in Fig.~\ref{F1}(a) clearly confirm anti-parallel spin alignment between the interstitial and nodal spins and the parallel alignment of interstitial spins within the classical FRI phase, whereas an increase in temperature may be responsible for an up-rise of small transverse (quantum) correlations. It is clear from Fig.~\ref{F1}(b) that the longitudinal correlation between the interstitial spins $\left<S_{i1}^z S_{i2}^z\right>$ is 
antiferromagnetic at low but ferromagnetic at high temperatures, 
which indicates a spin frustration above the UPA ground state according to the concept of temperature-dependent spin frustration \cite{frustr}. The spin correlations within the interstitial spins do not reach unity due to the presence of strong quantum fluctuations associated with the hopping process. The complete spin alignment of the interstitial and nodal spins relevant to the SPA ground state is evident from the correlation functions depicted in Fig.~\ref{F1}(c), where a small increase in transverse correlation function induced by thermal fluctuations can be also observed. Finally, Fig.~\ref{F1}(d) illustrates the correlation functions exactly at the highly degenerate triple point, at which zero temperature asymptotic limits reach nontrivial values $\left<S_{i1}^{x,y} S_{i2}^{x,y}\right>=(5-\sqrt{5})/20$,  $\left<S_{i1}^z S_{i2}^z\right>=(5+3\sqrt{5})/20$ and $\left<\sigma_{i1}^z S_{i1}^z\right>=(3\sqrt{5}-5)/10$ (see Appendix~A).

\begin{figure}[htbp]
\centering
\includegraphics [scale=0.4,clip]{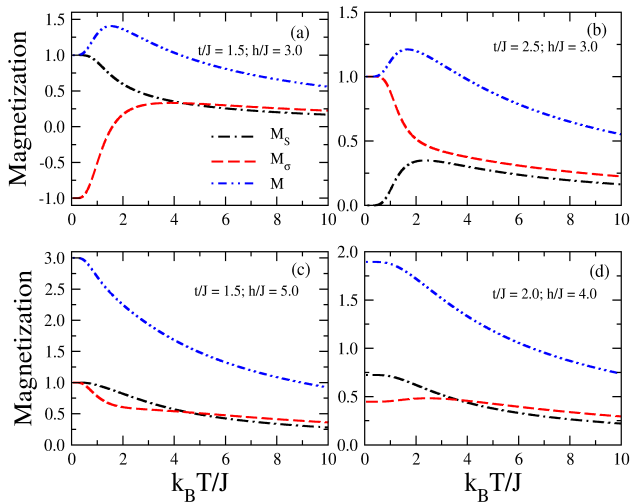}
\caption{Temperature dependence of the single-site magnetization $M_S$ of the interstitial spins, the single-site magnetization $M_\sigma$ of the nodal spins and  the total magnetization ($M$) per  diamond unit for the set of parameters: (a) $t/J=1.5$, $h/J=3.0$ (FRI phase); (b) $t/J=2.5$, $h/J=3.0$ (UPA phase); (c) $t/J=1.5$, $h/J=5.0$ (SPA phase); (d) $t/J=2.0$, $h/J=4.0$ (triple point).}
\label{F2}
\end{figure}

Thermal behavior of the total magnetization $M$ per diamond unit (dash-dot-dot lines) is plotted in Fig.~\ref{F2} along with the single-site magnetization $M_S$ of the interstitial spins (dash-dot lines) and the single-site magnetization $M_\sigma$ of the nodal spins (dash lines). Fig.~\ref{F2}(a) serves in evidence that the rising temperature may facilitate a spin reversal of the nodal spins within the FRI phase, which in turn causes a unusual thermally-induced increase in the total magnetization. Thermal excitations from the UPA ground state may also give rise to a striking increase of the total magnetization, which occurs due to a spin reorientation of the interstitial rather than the nodal spins (see Fig.~\ref{F2}(b)). Contrary to this, the magnetization of the nodal as well as interstitial spins show only a smooth monotonous decrease with the increasing temperature when starting from the SPA ground state (see Fig.~\ref{F2}(c)). Last but not least, the magnetization curves for the highly degenerate 
triple point is illustrated in Fig.~\ref{F2}(d). These  start from the nontrivial zero-temperature asymptotic values $M_{S}=(5+\sqrt{5})/10$, $M_{\sigma}=\sqrt{5}/5$ and $M=(5+2\sqrt{5})/5$ (see Appendix~A).

\begin{figure}[htbp]
\centering
\includegraphics[scale=0.44,clip]{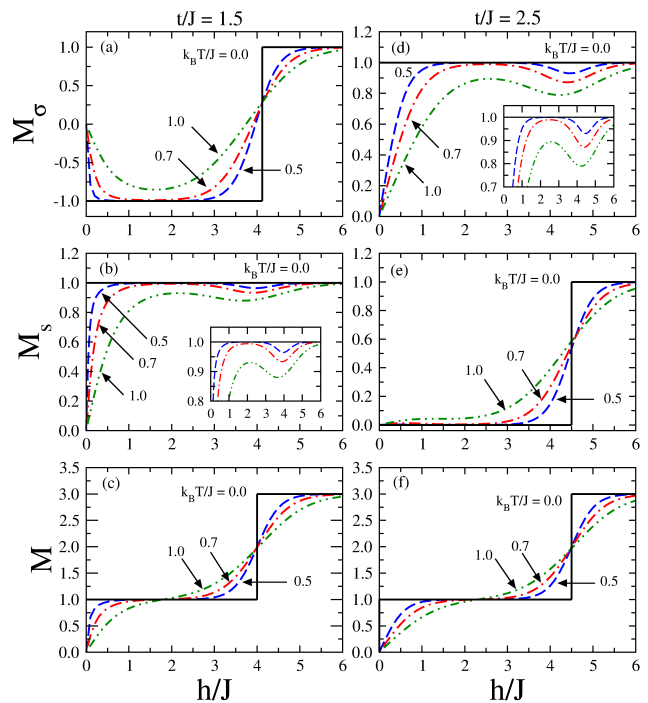}
\caption{The field dependence of the  magnetization for a few different values of the temperature and two selected values of the hopping term $t/|J|=1.5$ and $2.5$. The upper panel shows the single-site magnetization of the nodal spins, the central panel the single-site magnetization of the interstitial spins and the lower panel the total magnetization per diamond unit.}
\label{F21}
\end{figure}

Both available magnetization scenarios are displayed in Fig. \ref{F21}. The magnetization curves shown in Figs. \ref{F21}(a)-(c) shed light on a field-induced transition from the FRI phase to the SPA state, which occurs just for a sufficiently small values of the hopping term $t/J<2$. Strictly speaking, the true discontinuous phase transition associated with the magnetization jump of the nodal spins occurs at zero temperature only (see Fig. \ref{F21}(a)). However, it is worthwhile to remark that the tendency of the nodal spins to align in opposite to the magnetic field is clearly visible at non-zero temperatures as well. A sudden reorientation of the nodal spins in a vicinity of the saturation field causes a subtle decline of the magnetization of the interstitial spins due to the antiferromagnetic coupling between the nodal and interstitial spins (see the inset in Fig. \ref{F21}(b)). Another magnetization scenario shown in Figs. \ref{F21}(d)-(f) relates to a discontinuous field-induced transition from the 
UPA state to the SPA state. It can be clearly seen from Fig. \ref{F21}(e) that the observed magnetization jump near the saturation field occurs on account of a spin reversal of the interstitial spins. The sudden change in the magnetization of the interstitial spins now evokes a subtle decline of the magnetization of the nodal spins. Altogether, it could be concluded that the magnetization curve always displays a magnetization plateau at one-third of the saturation magnetization even though the microscopic mechanism for a plateau formation may be different.

\subsection{Spin frustration}

It is quite apparent from the eigenvector (\ref{34}) of the UPA ground state that the quantum-mechanical motion of two mobile electrons on the interstitial sites gives rise to a kinetically-induced spin frustration of the nodal Ising spins at zero temperature. To verify a frustrated character of the nodal spins at finite temperatures one may take advantage of the concept of temperature-dependent frustration, which requires a negative sign for the product of correlation functions along an elementary plaquette implying incapability of the spins to satisfy all underlying spin-spin interactions. The negative sign of the product $\left<S_{i1}^z S_{i2}^z\right>\left<\sigma_{i1}^z S_{i1}^z\right>^2$ can be achieved just if the longitudinal correlation function between the interstitial spins $\left<S_{i1}^z S_{i2}^z\right>$ is predominantly antiferromagnetic (negative). It has been demonstrated in the previous section that the correlation function $\left<S_{i1}^z S_{i2}^z\right>$ indeed changes its sign at a certain 
temperature when starting from the UPA ground state, whereas the relevant frustration temperature can be regarded as an indicator of the transition from the frustrated regime to the non-frustrated one.

\begin{figure}[htbp]
\centering
\includegraphics [scale=0.36,clip]{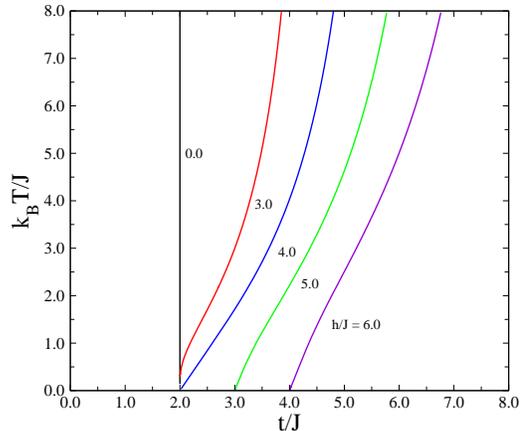}
\caption{The frustration temperature as a function of the hopping term for a few selected values of the magnetic field.
The investigated system is frustrated below the displayed lines where the product of the longitudinal spin correlations around a plaquette is negative. Frustration is only present above the UPA ground state.}
\label{F3}
\end{figure}

The frustration temperature, as calculated from crossing points of the longitudinal correlation function between the interstitial spins $\left<S_{i1}^z S_{i2}^z\right>$, is plotted in Fig.~\ref{F3} against the hopping term $t/J$ for a few selected values of the magnetic field. It turns out that the antiferromagnetic correlation $\left<S_{i1}^z S_{i2}^z\right> < 0$ develops just for a sufficiently strong hopping term $t/J > 2$, because the coupled spin-electron diamond chain may become frustrated only within the UPA phase. The frustration temperature accordingly starts at low enough magnetic fields
$h/J \leq 4$ from the ground-state boundary between the FRI and UPA phases at $t/J=2$, while the initial value of the frustration temperature
is progressively shifted along the ground-state boundary between the SPA and UPA phases $t/J=h/J-2$ at higher magnetic fields $h/J > 4$. Altogether, it can be concluded that the frustrated regime of the correlated spin-electron diamond chain is delimited from above by the displayed curves of frustration temperature.

\subsection{Fermionic concurrence}

\begin{figure}[htbp]
\centering
\includegraphics [scale=0.36,clip]{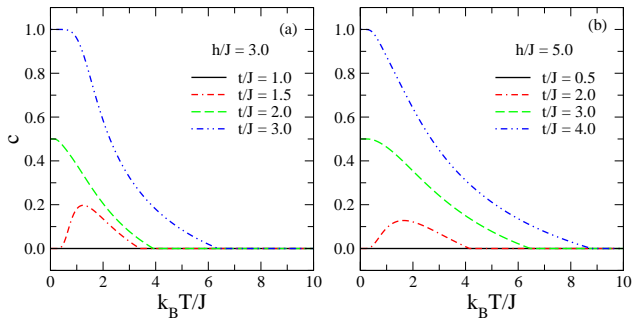}
\caption{The fermionic concurrence versus temperature for several values of the hopping term and two different values of the magnetic field: (a) $h/J=3.0$ (transition from the FRI to the UPA ground state); (b) $h/J=5.0$ (transition from the SPA to the UPA ground state). The re-entrant behavior emerges when the hopping amplitude drives the investigated system to the classical FRI or SPA ground state, but still keeps it in a close vicinity of the ground-state boundary with the quantum UPA ground state.}
\label{C1}
\end{figure}

Last but not least, we will turn our attention to a discussion of the bipartite quantum entanglement between two mobile electrons located at the interstitial sites of a given plaquette, which will be quantified by the fermionic concurrence calculated according to Eq. (\ref{Concur}). The temperature dependence of the fermionic concurrence is depicted in Fig.~\ref{C1} for several values of the hopping term and two different values of the magnetic field. As it could be expected, the concurrence is zero well within the classical FRI phase as illustrated by the solid line corresponding to $t/J=1.0$ in Fig.~\ref{C1}(a). On the other hand, the fermionic concurrence within the fully entangled UPA ground state for $t/J=3.0$ (dash-dot-dot line) starts from its maximum value and then gradually decreases with increasing temperatures. The special case with the fixed value of the hopping term $t/J=2.0$ corresponds to the phase coexistence of the FRI and UPA phases at zero temperature (the dashed line) and hence, the zero-
temperature limit of the fermionic concurrence is given by the mean value of the disentangled FRI phase and the fully entangled UPA phase. However, the most striking finding concerns with a re-entrant behavior of the fermionic concurrence when the hopping parameter is sufficiently close but slightly below the ground-state boundary between the FRI and UPA phases. Under this condition, the fermionic concurrence starts from zero in agreement with the classical character of the FRI ground state, then it emerges at some lower threshold temperature due to thermal excitations to the UPA  phase and finally, it completely disappears at a certain higher threshold temperature (see the dash-dot line for the fixed value $t/J=1.5$) .

It can be understood from Fig.~\ref{C1}(b) that the fermionic concurrence displays qualitatively similar thermal variations also at higher magnetic fields when the variation of the hopping term causes a phase transition between the UPA and SPA ground states. As one can see, the concurrence always starts from zero in accordance with the classical character of the SPA ground state for small enough hopping terms, but it may eventually show the interesting re-entrant behavior when the hopping term is selected sufficiently close to the ground-state boundary between the SPA and UPA phases (see dash-dot line for the hopping term $t/J=2.0$ in Fig.~\ref{C1}(b)). The dashed line shows the fermionic concurrence exactly at the ground-state boundary between the SPA and UPA phases ($t/J=3.0$ for $h/J = 5.0$) when the concurrence starts from its mean value $C = 1/2$. If the hopping term is strong enough to enforce the UPA ground state, then, the fermionic concurrence starts from its maximum value and afterward gradually 
diminishes with the rising of temperature until it completely vanishes above a certain threshold temperature.

\begin{figure}[htbp]
\centering
\includegraphics [scale=0.32,clip]{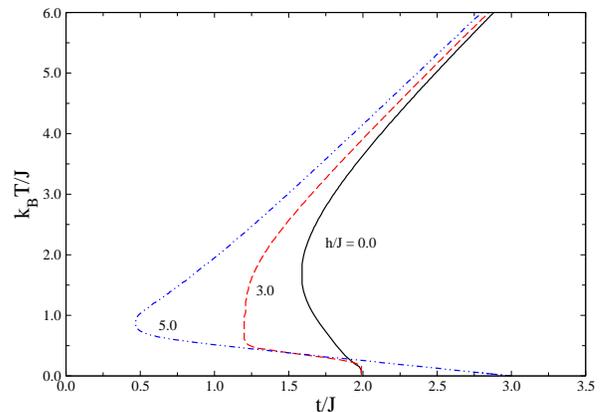}
\caption{The threshold temperature as a function of the hopping amplitude for three different values of the magnetic field: $h/J=0.0$, $3.0$ and $5.0$. The system is entangled (disentangled) inside (outside) the area delimited by the displayed lines. The re-entrant behavior indicates that thermal fluctuations favors quantum entanglement in the close vicinity of the FRI-UPA and SPA-UPA ground-state phase boundaries.}
\label{C2}
\end{figure}

The threshold temperature for the disappearance of the fermionic concurrence is plotted against the hopping term in Fig.~\ref{C2} for three different values of the magnetic field. The mobile electrons on the interstitial sites are fully entangled just within the UPA phase and hence, it follows that all depicted lines of threshold temperature should start from the ground state boundary between the UPA-FRI and UPA-SPA phases, respectively. The dependences  presented in Fig.~\ref{C2} are indeed consistent with this statement and they also provide an independent confirmation of the striking re-entrance when the quantum entanglement is induced above the classical FRI or SPA phase on account of vigorous thermal excitations to the UPA state.

\begin{figure}[htbp]
\centering
\includegraphics [scale=0.35,clip]{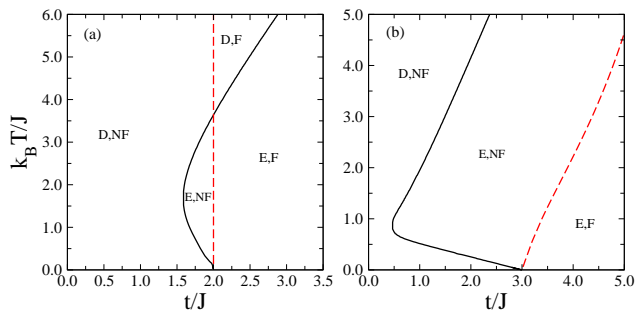}
\caption{A comparison between the threshold and frustration temperatures plotted as a function of the hopping amplitude for two different magnetic fields: (a) $h/J=0.0$; (b) $h/J=5.0$. The notation for parameter regions are: Entangled (E), Frustrated (F), Disentangled (D), and Non-frustrated (NF). The re-entrance in the threshold temperature depicted in Fig. \ref{C3}(a) implies the emergence of a low-temperature entangled-non-frustrated regime, as well as, a high-temperature disentangled-frustrated regime.}
\label{C3}
\end{figure}

At this stage, it might be quite interesting to study a mutual correlation between the spin frustration and quantum entanglement, because the kinetically-driven spin frustration of the nodal spins appears due to the hopping process of the mobile electrons with opposite spins inevitably underlying the quantum entanglement. Bearing this in mind, the fermionic quantum entanglement could be thus regarded as an indispensable ground for the kinetically-driven frustration of the nodal spins. It actually turns out that the zero-temperature asymptotic value of the threshold and frustration temperature always start from the same value of the hopping term, whereas the threshold temperature for the quantum entanglement generally lies slightly above the frustration temperature at low and moderate temperatures. However, the threshold and frustration temperature might also cross each other at a certain higher temperature as illustrated in Fig. \ref{C3}(a), which gives evidence that the spin frustration of nodal spins may 
emerge at higher temperature despite of disentangled character of the interstitial spins. However, we must have in mind that the crossing of the frustration and  threshold temperature takes place at a certain temperature, which is however high enough to produce only very weak spin-spin correlations around the elementary plaquette.

\section{Summary and Conclusions}

In the present work, we have provided a detailed analysis of the relationship between spin frustration and quantum entanglement in a hybrid spin-electron system containing the localized Ising spins and mobile electrons. In particular, we have considered the diamond chain prototype model with the localized Ising spins situated at the nodal lattice sites that interact with two delocalized electrons quantum-mechanically hopping between a pair of the interstitial sites. The model was exactly solved using the transfer-matrix technique. The ground-state phase diagram was shown to be composed of two classical SPA and FRI phases in addition to the one quantum UPA phase, in which the interstitial electrons are quantum-mechanically entangled. The relevant spin-spin correlations around an elementary plaquette were obtained and unveiled a low-temperature frustrated regime in the parameter region corresponding to the UPA ground state. In this regime there is no local spin ordering that satisfies simultaneously all signs 
of the spin correlations. In the present model frustration results from the quantum hopping process that favors the anti-parallel spin alignment of the interstitial electrons. We have also analyzed the distinct magnetization processes and showed the occurrence of intermediate plateau at one-third of the saturation magnetization. Also, we have found that the spin re-orientation of one sub-lattice (nodal or interstitial spins) observable in a vicinity of the field-induced transition is responsible for a small drop in the magnetization of the complimentary sub-lattice. Finally, we have quantified the degree of fermionic entanglement between two interstitial electrons by computing the quantum concurrence in the spin up sub-space. 

The low-temperature asymptotic value of the fermionic concurrence has unambiguously confirmed an existence of the quantum entanglement within the UPA ground state, while it has also demonstrated rather unexpected re-entrant behavior due to thermally activated excitations when the hopping term drives the investigated model towards the disentangled FRI and SPA ground states but still preserves it in  a vicinity of the ground-state boundary with the quantum UPA  ground state. We have also compared the threshold and frustration temperatures with the goal to bring insight into a mutual relationship between the quantum entanglement and geometric spin frustration. It has been convincingly evidenced that these two physical properties are strongly related yet independent. Besides the conventional entangled-frustrated and disentangled-non-frustrated regimes, we have additionally found entangled-non-frustrated and disentangled-frustrated regimes as well. Although geometric spin frustration and quantum entanglement are 
closely related features within the present model, this result indicates that thermal fluctuations have distinct influence on these two physical aspects. It would be valuable to have further studies of quantum spin chains with competing interactions to shed additional light of the relevant thermal processes influencing the interplay between spin frustration and quantum entanglement.

\section{Acknowledgments}

This work was partially supported by CNPq, CAPES, FINEP (Brazilian Research agencies), as well as by FAPEAL (Alagoas State Research agency). 
J.S. acknowledges warm hospitality during his stay at Universidade Federal de Alagoas supported under Science Without Borders programme.

\section{Appendix A}
\begin{figure}[htbp]
\centering
\includegraphics [scale=0.4,clip]{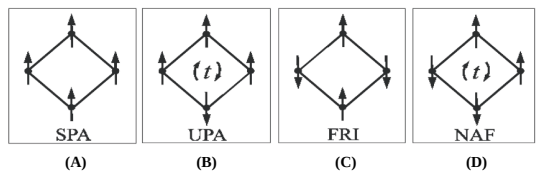}
\caption{Degenerated plaquette configurations at the co-existence point with the coordinates $h/J=4$ and $t/J=2$: (a) daturated paramagnetic state (SPA) with the interstitial and nodal spins parallel to the external field; (b) unsaturated paramagnetic state (UPA) with nodal spins parallel to the external field and interstitial electrons antiparallel to each other in an entangled state; (c) ferrimagnetic state (FRI) with the nodal spins anti-parallel to the field and interstitial spins parallel to the external field; (d) nodal antiferromagnetic state (NAF) with the nodal spins anti-parallel with respect to each other and interstitial spins parallel to the external field.}
\label{AA1}
\end{figure}
At the point of co-existence between the SPA, FRI and UPA phases, their corresponding configurations in a plaquette  are also degenerated with a nodal antiferromagnetic phase (NAF). Therefore, in the ground state, cells with either one of these configurations are allowed (see Fig.~10). However, NAF cells can only appear in pairs and  only FRI cells can be placed between them. Defining $N_{SPA}$, $N_{UPA}$, $N_{FRI}$, and $N_{NAF}$ as the number of cells in each configuration ($N=N_{SPA}+N_{UPA}+N_{FRI}+N_{NAF}$) the number of possible ways to distribute then along the chain can be written as
\begin{eqnarray}
\Omega &=& \frac{(N_{SPA}+N_{UPA})!}{N_{SPA}!N_{UPA}!}\times \nonumber \\ 
~&~& \frac{\left(N_{SPA}+N_{UPA}+N_{NAF}/2\right)!}{(N_{SPA}+N_{UPA})!\left(N_{NAF}/2\right)!}  \times \nonumber \\
~&~& \frac{\left(N_{NAF}/2+N_{FRI}\right)!}{\left(N_{NAF}/2\right)!N_{FRI}!}, \label{AAA1}
\end{eqnarray}
where the first fraction accounts for the distinct permutations between $SPA$ and $UPA$ cells, the second fraction accounts for the number of ways NAF dimers and paramagnetic cells (either SPA or UPA) can be arranged, and the last one corresponds to the number of possible ways FRI cells can be distributed between NAF dimers. The actual fraction of cells in each configuration is found by maximizing $\Omega$, resulting in 
\begin{eqnarray}
N_{SPA}/N&=&\frac{5-\sqrt{5}}{10}, \\
N_{UPA}/N&=&\frac{5-\sqrt{5}}{10}, \\
N_{FRI}/N&=&\frac{10-4\sqrt{5}}{10}, \\
N_{NAF}/N&=&\frac{-10+6\sqrt{5}}{10}.
\label{AA3}
\end{eqnarray}
The values of magnetization of the  nodal and interstitial sites, as well as the relevant correlation functions in each configuration are summarized in table I. The proper average of these quantities, weighted considering the relative fraction of cells in each configuration, provides the values reported in the main text at the triple co-existence point $h/J=4$ and $t/J=2$.

\begin{table}[h!]
\label{tableone}
\caption{Fraction of cells $N_X/N$ (X=SPA, UPA, FRI, NAF) in each  cell configuration  at the co-existence point $(h/J=4$, $t/J=2)$. The corresponding values of the nodal magnetization $M_{\sigma}$, interstitial magnetization $M_S$, correlation between nodal and interstitial spins $\langle \sigma_i^zS_{i1}^z \rangle$, correlation between the spin $z$-components of the interstitial sites $\langle S_{i1}^zS_{i2}^z \rangle$, and between their spin x-components $\langle S_{i1}^xS_{i2}^x \rangle$ are also shown. Their weighted average provides the values reported in the main text. }
\begin{ruledtabular}
\begin{tabular}{ccccccc}
Plaquette & $N_X/N$ & $M_{\sigma}$ & $M_S$ & $\langle \sigma_i^zS_{i1}^z \rangle$ & $\langle S_{i1}^zS_{i2}^z \rangle$ & $\langle S_{i1}^xS_{i2}^x \rangle$\\~\\
SPA&$\frac{5-\sqrt{5}}{10}$& 1 & 1 &1&1&0\\
UPA&$\frac{5-\sqrt{5}}{10}$& 1 & 0 &0&-1/2&1/2\\
FRI&$\frac{10-4\sqrt{5}}{10}$& -1 & 1 &-1&1&0\\
NAF&$\frac{-10+6\sqrt{5}}{10}$& 0 & 1 &0&1&0\\
\end{tabular}
\end{ruledtabular}
\end{table} 

\section{Appendix B}

The relevant thermal averages of quantities associated with a pair of interstitial electrons can be obtained as a function of the proper reduced matrix. One starts by defining the auxiliar operator associated with the eigenstates of a pair of interstitial electrons \cite{rojas1}:
\begin{eqnarray}
\varrho(\sigma_i,\sigma_{i+1})=\sum_{j=1}^6 {\rm e}^{-\beta\varepsilon_{ij}(\sigma_i,\sigma_{i+1})}\left|\varphi_{ij}\right>\left<\varphi_{ij}\right|.
\label{AB1}
\end{eqnarray}
which depends on the spin orientation of the neighboring Ising spins. The reduced density matrix elements can be obtained by tracing out all Ising spins and all interstitial electrons except those at a given plaquette. It can be directly shown that the reduced density matrix elements can be put into the form:
\begin{eqnarray}
\rho_{k,l}\!\!\!&=&\!\!\! \frac{1}{\mathcal{Z}_N }\sum_{\{\sigma\}}(\omega(\sigma_1,\sigma_2)\ldots\omega(\sigma_{r-1},\sigma_r)\varrho_{k,l}(\sigma_r+\sigma_{r+1})\nonumber \\
        \!\!\!&\times&\!\!\! \omega(\sigma_{r+1},\sigma_{r+2})\ldots\omega(\sigma_N,\sigma_1))\! \nonumber \\
\rho_{k,l}\!\!\!&=&\!\!\!\frac{1}{\mathcal{Z}_N}\rm{Tr}\left(\textbf{P}_{k,l}{W}^{N-1}\right),
\label{AB3}
\end{eqnarray}
where we introduced the matrix
\begin{eqnarray}
\textbf{P}_{k,l}=\left( \begin{array}{clrr}
\varrho_{k,l}(2) & \varrho_{k,l}(0)\\
\varrho_{k,l}(0) & \varrho_{k,l}(-2)\\
\end{array}\right) \label{AB4}.
\end{eqnarray}
and the indices $k$ and $l$ run from 1 to 6 according to the possible states of the interstitial sites  $\left|\uparrow,\uparrow\right>_i$, $\left|\downarrow,\downarrow\right>_i$, $\left|\uparrow\downarrow,0\right>_i$, $\left|\uparrow,\downarrow\right>_i$, $\left|0,\uparrow\downarrow\right>_i$ and $\left|\downarrow,\uparrow\right>_i$.

By exploring the unitary transformation that diagonalizes the transfer matrix $\omega$ (see main text), the density matrix elements result in  
\begin{multline}
\rho_{k,l}= \frac{\varrho_{k,l}(2)+\varrho_{k,l}(-2)}{2\lambda_+}+\frac{2\varrho_{k,l}(0)\omega(0)}{Q\lambda_+}    \\
+\frac{\left[\varrho_{k,l}(2)-\varrho_{k,l}(-2)\right]\left[\omega(2)-\omega(-2)\right]}{2Q\lambda_+}.
\label{AB6}
\end{multline}

As a function of the reduced density matrix elements, one can readly obtain the thermodynamic averages
\begin{eqnarray}
\left<c_{i1,\uparrow}^\dag,c_{i2,\uparrow}\right>&=&2\rho_{5,4} \\ \nonumber
\left<n_{i1,\uparrow},n_{i2,\uparrow}\right>&=&\rho_{1,1}\\ \nonumber
\left<n_{i1,\uparrow}\right>&=&\rho_{1,1}+2\rho_{3,3} \nonumber
\label{AB8}
\end{eqnarray}
where the identities $\rho_{5,4}=\rho_{6,3}$ and $\rho_{3,3}=\rho_{4,4}$ were used. From the above expressions, the quantum concurrence can be explicitly written as
\begin{eqnarray}
\mathcal{C}=\rm{max}\left\{0, 4\rho_{5,4}-2\sqrt{\rho_{1,1}(1-\rho_{1,1}-4\rho_{3,3})}\right\}
\label{AB7}
\end{eqnarray}

\end{document}